\input epsf
\documentstyle[preprint,prl,aps]{revtex}
\tightenlines

\begin{document}

\date{January 7, 1999}
\draft
\title{JACOBI EVOLUTION OF STRUCTURE FUNCTIONS: \\
CONVERGENCE AND STABILITY}
\author{G.Shaw}
\address{Theoretical Physics Group, Department of Physics and Astronomy, \\
The University of Manchester, M13 9PL, UK}
\maketitle

\begin{abstract}
The Jacobi evolution method has been widely used in the  QCD 
analysis of structure function data. However a recent paper claims that there
are serious problems with its convergence and stability. Here we briefly
review the evidence for the adequate convergence of the method; and show 
that there are errors in the above paper which undermine its conclusions.

\end{abstract}

\section{INTRODUCTION}

In analysing data on nucleon structure functions, it is convenient to
have as fast and simple a method of implementing their QCD evolution 
as possible. For this reason, the Jacobi evolution 
method \cite{PS,BLS,Russians} has been adopted in many 
recent analyses\footnote{A SPIRES search reveals more than 30 citations since 
1995.}, including 
applications to polarized \cite{Leader} and 
non-forward or off-diagonal \cite{Belitsky} parton distributions. 
However,  Ghosh and Raha \cite{GR} have made the startling 
claim that it is unstable due to convergence problems.
In this paper, we discuss the convergence properties of the method in general,
and the problem raised by Ghosh and Raha in particular. 


We start by very briefly summarizing the method \cite{PS,BLS,Russians}. The 
basic idea is to expand a given structure function 
 $\Delta(x,Q^2)$ at fixed $Q^2$ in the form
\begin{equation}
\Delta(x,Q^2) = x^\beta (1-x)^\alpha \sum_{k=0}^{\infty} a_k(Q^2) 
\Theta_k^{\alpha \beta}(x)  \; ,
\label{i0} 
\end{equation}
where $\alpha, \; \beta > -1$ are real numbers  and 
$\Theta_k^{\alpha \beta}(x)$ 
are a complete set of Jacobi polynomials of order $k = 0,1,2,\ldots$. They are
 defined 
to satisfy\footnote{We follow the standard notation and conventions of
\cite{BLS}, where further mathematical details  may  be 
found.}
\begin{equation}
\int_0^1 dx \, x^\beta (1-x)^\alpha \Theta_k^{\alpha \beta}(x) \,
\Theta_l^{\alpha \beta}(x) = \delta_{kl} \; \; ,
\label{i2}
\end{equation}
so that  the ``Jacobi moments'' 
\begin{equation}
a_k(Q^2) = \int_0^1 dx \, \Delta(x,Q^2)\,\Theta_k^{\alpha \beta}(x) \; . 
\label{i3}
\end{equation}
In practice, the truncated expansion
\begin{equation}
\Delta(x,Q^2) = \Delta_N(x,Q^2) \equiv 
 x^\beta (1-x)^\alpha \sum_{k=0}^{N} a_k(Q^2) 
\Theta_k^{\alpha \beta}(x) 
\label{i8} 
\end{equation}
is used, where $N$ is hopefully large enough for this to be a good 
approximation. The  structure function can then be evolved in  $Q^2$ 
by using the appropriate QCD evolution equations for the ``Jacobi moments.''
 For non-singlet structure functions
these  are of the form 
\begin{equation}
a_k(Q^2) =  \sum_{j=0}^{k} E_{jk}^{\alpha, \beta}(Q^2, Q_0^2, \Lambda) 
a_j(Q_0^2) \; ,  
\label{i7}
\end{equation} 
where  the $E_{jk}^{\alpha, \beta}(Q^2, Q_0^2, \Lambda)$ are known 
coefficients 
and $Q_0^2$ is some conveniently chosen reference $Q^2$.
Alternatively and equivalently, the  Jacobi moments $a_k(Q^2)$ can be
expressed in terms of the more familiar Cornwall-Norton moments at the 
reference $Q_0^2$.  For singlet structure functions, the evolution equations 
are
modified by the addition of appropriate inhomogeneous terms dependent on
the moments of the gluon  distributions at the reference $Q^2_0$.

\section{CONVERGENCE OF THE JACOBI EXPANSION}

The conditions for uniform convergence \cite{Russians,Szego} of the expansion 
 (\ref{i0}) in the range $0 < x < 1$ are the existence of the integrals
\begin{equation}
\int_{0}^{1}dx \,  x^\beta (1-x)^\alpha \Delta(x,Q^2) \hspace{1cm}
\int_{0}^{1} dx \, x^\beta (1-x)^\alpha \left[ 
\frac{\Delta(x,Q^2) - \Delta(z,Q^2)}{x-z} \right]^2
\label{cs1} 
\end{equation}
for $ 0 < z  < 1$.  For non-singlet structure functions, assumed
to be  analytic and bounded, these are satisfied for the whole range of
values $\alpha, \beta > -1$ for which Jacobi polynomials exist.
For singlet structure functions, assumed to behave like $x^{- \gamma}$
as $ x \rightarrow 0$ with $\gamma \ge 0$, they are satisfied for
$\alpha >  -1$, $\beta > -1 + 2 \gamma$.
The key question is not whether a convergent series exists, but whether it 
converges
sufficiently rapidly for the truncated expansion (\ref{i8}) to be a good
approximation for suitably chosen values of $\alpha$ and $ \beta$. These 
values can reflect the observed behaviour at small and large $x$,
or simply be varied  in order to improve the convergence.

The accuracy of the truncated expansion (\ref{i8}) can be checked in several
different ways. By construction  (\ref{i8},\ref{i7}) automatically 
guarantees the correct evolution of the first $N$ Cornwall-Norton moments, 
but other  QCD evolution properties
will only be correctedly reproduced if the series (\ref{i8})has converged. 
For example,   integrals of the form 
\begin{equation}
\int_0^1 dx \, \Delta(x, Q^2) \, / \, x  
\label{cs2}
\end{equation}
are sensitive to the low-$x$ region,  and  correspond to the well-known
Gross-Llewellyn-Smith and Gottfried sum rule integrals for $\Delta  = x F_3$ 
and  $ \Delta = F_2^{ep} - F_2^{en} $  respectively. These 
are independent of $Q^2$ in LO QCD, but this  is only guaranteed 
by (\ref{i8}) if the series has converged. Barker, Langensiepen and
Shaw \cite{BLS}  checked  
this property in their LO fits  and found that 
with $N=10$ it was  satisfied over the range $ 2 \le Q^2 \le 200$ GeV$^2$ 
to within an  accuracy of $10^{-3}$. 
 They also checked that the QCD evolution of the higher
Cornwall-Norton moments $M_i(Q^2)$ for $i=15,20$ was accurately 
reproduced\footnote{In LO,
the logarithm of the Cornwall-Norton moments depends linearly on
$ t  \ln \left[  \ln \,Q^2/\Lambda^2 \right] \; .$
This linear dependence was reproduced with a fitted slope within one per
cent of the QCD prediction.} and  concluded that 
truncating at   $N = 10$ was sufficient for all practical purposes over the 
kinematic  range considered.

Other methods are to check  the variation of the predictions 
of (\ref{i8}, \ref{i7}) with increasing $N$; and to compare them with the 
results of ``exact'' 
numerical evolution of the DGLAP equations for the same input.  Both 
methods were exploited for non-singlet strucure functions by  
Chyla and Rames \cite{CR}, who compared the results of ``exact'' QCD evolution
with the results obtained by trucating the Jacobi series at $ N$ values 
varying up to 20. Specifically, they assumed the input form
\begin{equation}
 \Delta(x, Q^2) = 4.05 \, x^{0.404}(1-x)^{3.73}  \hspace{1cm} (Q_0^2 = 90) 
\label{cs2a}
\end{equation}
and considered its evolution over  the
$Q^2$  range $ 1 \le Q^2 \le 1000$ GeV$^2$. 
They found a rapidly convergent  oscillatory 
approach to the ``exact solution'' using the  same 
values $\alpha = 3$, $\beta = 1/2$ as \cite{BLS}; and that small variations in
these values had little effect on the convergence. For example, detailed 
results are presented for the reasonable choice $\Lambda_{LO} = 0.1$,
retaining terms up to $N=9$ . Even   at 
 $x = 0.7$ where the structure function is  very 
small, it is reproduced with an accuracy of $\le 1 \%$,
corresponding to an absolute error of $\le 0.0005$ This is very small on
the scale of variation of the structure function itself, which has a maximum 
value of order unity. 

Chyla and Rames \cite{CR} also made similar studies for the other
 evolution methods which had been  proposed and  concluded 
that the Jacobi  method  
was {\em ``by far the simplest, fastest and simultaneously very accurate way of
analysing the $Q^2$ evolution of structure functions.''} 

Finally, we note that for singlet structure functions, Krivokhizhin et 
al.\cite{Russians} have presented results for  the errors obtained
in reproducing a ``typical structure function'' 
$$
\Delta(x) = 2.67 \, x^{0.25}(1-x)^{3}+ 0.48(1-x)^8
$$ 
at fixed $Q^2$ using the truncated expansion (\ref{i8})  for $N$ values 
up to 17. For the most favourable values $\alpha =3, \; \beta = -0.85$
they found an average error $ < 10^{-4}$ for $N = 7$; and
for $N = 8$, the error at any $x$ was always less than $10^{-4}$.  
For other $\alpha, \beta$ values the convergence is less
rapid but still more than adequate for  practical purposes.

\section{ A POSSIBLE PROBLEM?}

Very recently, Ghosh 
and Raha \cite{GR} have again investigated the convergence  of
the Jacobi method  by assuming specific 
analytic forms for non-singlet structure functions at the 
reference $Q_0^2$. 
These are  approximated  by the  expansion (\ref{i8}) using (\ref{i3}) to 
evaluate the Jacobi moments at $Q_0^2$  and (\ref{i7}) to evolve them to
other $Q^2$. 

The basic problem is already manifest at the reference $Q^2 = Q_0^2$.
Plots are presented showing that as $ N$ 
is increased from zero, their computed
value of the truncated series $\Delta_N(x,Q_0^2)$ initially oscillates  
and then becomes
approximately   stable and equal to the exact $\Delta(x,Q_0^2)$ for a wide
range of $N$ values. This is similar to the behaviour found by other authors.
 However when $N$ increases further, 
beyond a  value  which is typically of order 20 - 25,
the results become  wildly unstable, presumably
reflecting large contributions from high order polynomials in the Jacobi 
series (\ref{i0}).
They therefore conclude that {\em ``the convergence breaks down 
completely for large $N$.''}

This reult is  surprising, since the input structure functions are
smoothly varying  and the series must eventually
converge uniformly, as noted above. However  the precise 
evaluation of both Jacobi polynomials and of the integrals (\ref{i3}) 
becomes non-trivial as the order increases; and no tests are reported with
a view to verifying the accuracy of the methods used for the highest orders
considered. The question therefore arises whether the
reported problem is really endemic to the Jacobi method; 
or whether it could arise from problems with the  procedures 
used, which are not specified in any detail. Fortunately, this can be 
easily settled using the first set of input forms considered, which  are of 
the generic type\footnote{Cf. eqns.(11) of \cite{GR}
for $\Delta = xu_v(x), \; xd_v(x)$}
\begin{equation}
\Delta(x,Q_0^2) = x^{3/2} (1-x^2)^3 \sum_{k=0}^{p} b_k \, (1 - x^2)^k \; .
\label{cs2b} 
\end{equation}
This can obviously be  rewritten  in the form
\begin{equation}
\Delta(x,Q_0^2)  = x^{1/2} (1-x)^3  \sum_{j=1}^{2p+4} c_j \,x^j \; .   
 = x^{1/2} (1-x)^3 \sum_{j=1}^{2p+4} a_j(Q_0^2) 
\Theta_j^{3,1/2}(x)  \; ,
\label{cs3} 
\end{equation}
where we have assumed the usual values $\alpha=3 \; \beta = 1/2 $
  for the Jacobi indices.
In other words, $a_k(Q_0^2) = 0$ 
for $ k > N_{max} = 2p + 4$
 and the Jacobi expansion,  implemented in the most straightforward 
way, terminates exactly after a small
number of terms. The problematic contributions found by Ghosh and 
Sitar\footnote{Another problem raised in their paper is that even if one 
restricts oneself to $N$ values in the range where the results are stable, 
their results are strongly dependent on whether they start their evolution
from $Q^2 = 3.5$ or $Q^2 = 5$ GeV$^2$.  This stems from two trivial
errors in the equations used to describe the input at $Q^2 = 3.5$. For 
full details
see the earlier and longer version of this paper, which is still accessible on
lanl hep-ph/9901253v1.}
at large $N$ are  completely excluded in this case. They are not endemic in
the Jacobi method,
and must arise  from some problem with their procedure, numerical or otherwise.

\section{SUMMARY}

Previous work has established that the Jacobi expansion coverges very
rapidly for reasonably chosen parameters $\alpha$, $\beta$.
In the light of the previous subsection, we do not believe 
that Ghosh and Raha have provided any reason to modify this conclusion.

\end{document}